# Outline of a novel architecture for cortical computation

Kaushik Majumdar, Institute of Mathematical Sciences, Chennai – 600113, India. E-mail: kaushik@imsc.res.in

**Abstract:** In this paper a novel architecture for cortical computation has been proposed. This architecture is composed of computing paths consisting of neurons and synapses. These paths have been decomposed into lateral, longitudinal and vertical components. Cortical computation has then been decomposed into lateral computation (LaC), longitudinal computation (LoC) and vertical computation (VeC). It has been shown that various loop structures in the cortical circuit play important roles in cortical computation as well as in memory storage and retrieval, keeping in conformity with the molecular basis of short and long term memory. A new learning scheme for the brain has also been proposed and how it is implemented within the proposed architecture has been explained. A few mathematical results about the architecture have been proposed, some of which are without proof.

## 1. Introduction

The term 'computation' has traditionally been pertinent to electronic computation only, which signifies the execution of algorithms within the realm of the classical information theory. To this date the most successful architecture as a framework for this type of computation has been the von Neumann architecture where the memory is accessed with hierarchical ease but always remains separated from the units of logical operations. The fastest accessible memory is stored in registrar or cache and called the working memory. On the contrary the general consensus among the neuroscientists is that the memory does not reside concentrated in a specific location in the brain, but rather the various forms of memories remain distributed over a wide part of the brain, where particular cortical regions like hippocampus and the limbic systems play crucial roles in mediating them [1]. Despite this well established paradigm for cortical computation von Neumann inspired architecture for cognitive computation [2] has been proposed [3]. In spite of the elegant logical structure of this model working memory cannot be treated as a unit separate from logical operations in the nervous systems (for a neural basis of memory see ref. [4]). A more realistic architecture for cortical computation called *dynamic core hypothesis* has been proposed in [5], [6]. According to this model varying subsets of sub-regions of the brain are assigned and unassigned during the recollection of conscious experiences. [3] looks mathematically more solid than neurobiologically sound. On the other hand [5] is biologically more solid than mathematically concrete. A more efficient architecture for cortical computation should fall somewhere in between these two. In this paper an elementary construction for such an architecture has been proposed. Biologically the model is so elementary that there is little scope for doubt about its validity. On the other hand it has been shown how deep mathematical investigations even this elementary model can warranty. Scope for nontrivial mathematical investigations to establish profound biological results have been outlined. Care has been taken never to deviate too far from the cellular and molecular neuroscience.

The importance of feed forward and feed back paths in neural information processing has been recognized through numerous studies. This paper has greatly been motivated by the survey presented in [7] and [19], the former is from a neurophysiological point of view and the latter from a computational point of view with considerable neuroanatomical insight. In case of the mammals information from the environment is collected through the sense organs on the body surface and transmitted through thalamus to the cortex where it gets processed hierarchically in various functional regions (olfactory senses do not reach cortex through thalamus [19]). This is called *feed forward* or *bottom up* processing. During this modulation of new information with old experience feed back signals are generated and propagated from higher processing regions to lower processing regions. This *top down* or *feed back* signal then starts monitoring the bottom up processing of information. In other words, past experience or learned knowledge takes an important part in mediating the feed forward processing. The whole processing then gets a structured form within the neural network of the brain in which brain regions get selectively involved. The information being processed can shift from one structure to another within the duration of execution of the task. This is the fundamental notion of dynamic core hypothesis. Note that the ability to shift from one structure to another endows a capability of combinatorial selection. This eventually enables a relatively small number of structures to process a vast body of different signals and produce very versatile outputs.

Apart from biological evidence existence of loops in the cortex can also be inferred mathematically. Cortex is a three dimensional structure. It has been proved in this paper that if the formation of an information processing line consisting of neurons and feed forward excitatory synapses is considered to be a random walk then the line will return to its initial position with a probability $0.2782$. This guarantees the formation of loop by a line with $\approx 28\%$ chance. From a theoretical point of view strong backward coupling was explored by Hopfield [9]. It has been argued in this paper that the loop structures play a central role in cortical information processing (also see [19], [20]).

Loops have been identified as the most basic system level computational units of the cortex [19], [20]. Multi-stage integration of processes going on in different cortical areas may be a common strategy throughout the cortex for producing complex behavior [25]. Loops traversing through different cortical areas and layers seem to be the most probable anatomical candidates to perform such integration in a self-sustaining manner. In the subsequent sections a simple but general neural circuit architecture as collection of loops will be described which resembles the circuit of the cortex. In section 2 the horizontal component of the architecture will be discussed. The vertical component will be taken care of in section 3. Some aspects of memory and learning under the light of the proposed architecture will be dealt in section 4. New scheme of cortical learning, called critical set learning, has also been proposed in this section. The paper will conclude with a section devoted to discussions and future directions.

## 2. Horizontal computation

In this paper cortical computation has been divided into lateral computation (LaC), longitudinal computation (LoC) and vertical computation (VeC). Among them LaC and LoC will constitute horizontal component of cortical computation and VeC the vertical component. The current section will be devoted to horizontal computation. LaC is the processing through the feed forward and feed back paths from lower brain regions (such as thalamus) to the higher brain regions (such as prefrontal cortex) and reverse respectively (described by the columns in Figure 1). LoC is the collection of feed forward and feed back processing taking place among the cortical regions and within a cortical region to locally influence the processing of information (described by the rows of Figure 1). There will usually be multiple lines (consisting of excitatory neurons pairwise joined with feed forward synapses) parallelly processing information from thalamus to prefrontal cortex. Several closely spaced parallel lines will make a *path*. The neurons belonging to various lines within the same path in a brain region may be interconnected with excitatory and inhibitory synapses directly and/or through local interneurons. Once the neurons in the LaC paths are activated the local LoC paths across those LaC paths are also activated. These LoC computations along the LaC paths have profound effect on the over all LaC computation. LoCs are not necessarily confined within a particular cortical region. They may span across several cortical regions. The input to the LoC circuits together with their anatomy and physiology mediate the LoC computation. The LaC activates the higher brain regions, which in turn sends feed back signals to the LoCs along the LaC paths. The LoCs then take part in mediating the LaC. This cycle of LaC driving LoC and LoC driving LaC to self sustain the cycle can keep alive the internal information processing within the brain long after the cessation of outside stimulus. This not only plays a fundamental role in consciousness and cognition but also in storage and retrieval of various forms of memory. Apart from loops both LaC and LoC consist open ended paths. However advantage of loops in neural information processing over the open ended paths has been shown in Lemmas 2.1 and 2.2 below. It is worthy to note that in a loop where all neurons are connected by feed forward excitatory synapses exciting any one nerve cell will activate the whole loop, provided each presynaptic neuron can make the postsynaptic neuron fire (this will be assumed to be valid in this paper).

**Definition 2.1:** A *big loop* will mean a cyclic brain circuit consisting of a longest feed forward path from the thalamus to the highest processing brain region and a feed back path from that region on or close to the thalamus (for an excellent review of the thalamocortical loops see [19]).

All input to the cortex, except for the olfactory sense, comes to it via thalamus [19] and therefore the big loops are the fundamental architectural units for the LaC. On the other hand LoC consists of smaller loop structures either within a region or spanning across several regions. If an area A projects to another area B then B also projects to A [20]. The existence of local collaterals is a major feature of the output pyramidal cells of the cortex. It allows the cortex to carry on local calculations indefinitely without further stimulation [20]. If those collaterals form cyclical connections (possibly through excitatory interneurons) then the following results will convince us about their ability to self sustain. The cycles of Lemma 2.2 below are LoC cycles. Lemmas 2.1, 2.2 and Theorem 3.1 have been taken from [10].

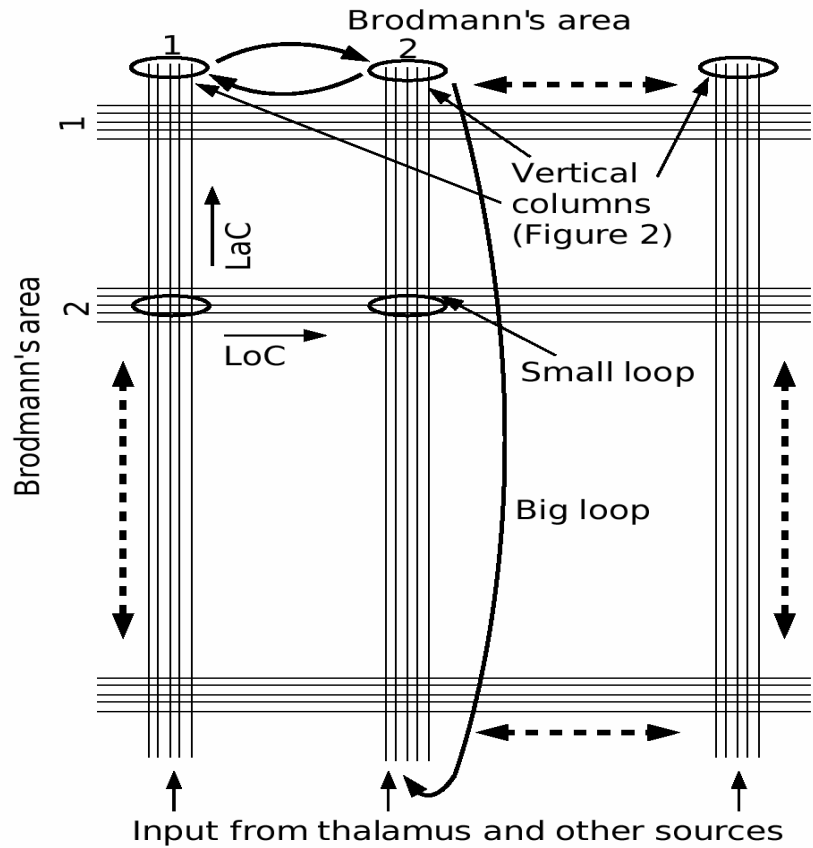

Figure 1. Horizontal computation consisting of lateral computation (LaC) and longitudinal computation (LoC). Big loops are components of LaC and small loops are components of LoC. The total number of Brodmann's areas has deliberately been kept blank to allow for flexibility. The horizontal and vertical lines signify dendrites and axons (please ignore the scale of the length). Synapses have not been shown. This figure should be viewed in conjunction with Figure 2 as described in subsection B.

**Lemma 2.1:** A cycle with $k$ nodes (neurons) in the directed graph of the brain circuit can be activated with greater probability than a linear path or line with the same number of nodes.

**Proof:** In a directed cycle if any neuron is activated the signal will propagate to all the other nodes cyclically and they will become activated in turn. Whereas to activate all the nodes in a line the first node must have to be activated. So if any neuron in the cycle can be activated with probability $p$ the whole cycle will be activated with probability $p$, whereas the whole line will have activation probability $\frac{p}{k}$ only. □

**Lemma 2.2:** Cycles in brain circuit can amplify signal.

**Proof:** Let a signal or an action potential of frequency $I$ (in case of a firing neuron frequency signifies intensity, so $I$ is also intensity of the signal) is reaching the $jth$ node of a cycle $(n_1,...,n_j,...,n_k)$, where $n_{k+1}=n_1$. If a signal takes time $T$ on an average to travel from one node to the next, and the incoming signal to $n_j$ is still reaching the node after $(k-1)T$ time, then the total dendritic input to $n_j$ becomes $I+I'$, where $I'$ has been received through the feed back loop. If $n_j$ is not already firing at the highest frequency the input $I+I'$ will make it fire at higher frequency than did $I$. □

**Theorem 2.1:** (*Contrast and amplify* principle)**:** During the bottom up journey signal travels from sensory neurons to the highest processing brain area by contrast and amplify principle.

**Proof:** Sensory inputs are carried by parallel LaC lines in the brain circuit from the thalamus and other input areas to the highest processing brain regions as shown in the schematic diagram of Figure 1. Any two of the parallel lines are short circuited by both excitatory and inhibitory interneurons. At the very beginning all the parallel lines carry signals from inputs according to the structure available at that instant, i.e., all the lines are active and between any pair of them excitatory and inhibitory interneurons are active or inactive without any control from the higher order brain regions. Once the signals reach in those regions 'an activity pattern emerges based on memory of past experience,' (this part has been explained following the proof) which is the *expected* pattern and it in turn starts controlling the interneurons (through feed back lines) by selectively short circuiting the pair wise parallel lines. A bunch of closely spaced lines short circuited mostly by excitatory interneurons work as a single path (in this theorem a path may be thought of as a collection of lines, see Figure 1) and two such groups of paths short circuited mostly by inhibitory interneurons work as different paths (not shown in the Figure 1). This way some information which are flowing through the same path are processed together as a cluster and two clusters become distinguished, because they are information carried by different paths. In the process the whole sensory input is decomposed by contrast.

Next, take a LaC path as a collection of parallel lines in the circuit of the brain circuit, which are pair wise short circuited mostly by excitatory interneurons. These short circuiting excitatory interneurons when form (directed) LoC cycles can enhance the signal passing through it according to Lemma 2.2. By the argument of Lemma 2.1 the signal enhancement will be more if $T$ is small, such as if most of the edges of the cycle are electrical synapses, $k$ is small and $I'$ is large (ideally close to $I$). □

**Question 2.1:** How 'an activity pattern emerges based on memory of past experience,' as claimed in the proof of Theorem 2.1?

**Answer:** Well, as a LaC path reaches a higher processing region like the hippocampal formation or the prefrontal cortex it gets access to neurons with more diverse connections and greater processing power. Let $A$ and $B$ be two LaC paths as collections of LaC lines

(recall that in Figure 1 each vertical line is a LaC line and their collection is a LaC path). $u$ be a neuron in the hippocampal formation connected to all LaC lines belonging to $A$ (presynaptically by dendritic arbor) and $B$ (postsynaptically by axonal branching) which will fire if at least $k$ presynaptic neurons fire. Now if sensory information coming through thalamus activates $k$ or more LaC paths of $A$ then the probability that $B$ will also be activated is $\sum_{j=k}^{r} \frac{r!}{j!(r-j)!} p_0^j (1-p_0)^{r-j}$, where $p_0$ is the probability of firing a presynaptic neuron to $u$ and $r$ is the number of activated LaC lines in $A$. To increase the chance of activating $B$ several $u$ will be needed. The higher is the number of $u$ the greater will be the success of activating $B$ through $A$. The pattern consisting of $A$ and $B$ connected through the class of neurons $u$ processes a given set of inputs in a specified way giving rise to a particular activity pattern, which is shaped by how $u$ neurons (to connect $A$ with $B$) have been recruited based on past experience. □

In the proof of the above claim $u$ may be treated as a class of neurons. Some cells in the class are most likely from the hippocampus. In this sense taking $p_0$ as a constant is an over simplification, although this is the predominant trend [11]. $p_0$ should ideally be a spatiotemporal function which can probably be modeled suitably by an appropriate wavelet function. That function in general will represent the spatiotemporal structure of synaptic plasticity in the areas of interest within the brain.

## 3. Vertical computation

Cortex is organized in six layers. Each cortical column of diameter about $0.5\,mm$ and height $2\,mm$ is the smallest computing unit of the cortex. In mammalian brain the neocortex is the region where most of the higher order neural computations take place. The neocortex receives input from the thalamus, from other cortical regions on both sides of the brain and from a variety of other sources. The output of the neocortex is also directed to several brain regions, including other regions of the neocortex on both sides of the brain, the basal ganglia, the thalamus, the pontine nuclei and the spinal cord. Different inputs to the neocortex appear to be processed in different ways and the outputs of the neocortex arise from different populations of neurons. The layering of neurons provides an efficient means of organizing the input-output relationships of the neocortical neurons [12]. The information flow within and among cortical columns has been schematically shown in Figure 2 (drawn after Figure 17-9 of [12], which signifies information flow in visual cortex). The cellular organization of cortical columns (which is not uniform across the cortex) has been described, e.g. in [12] (p. 327).

In the 3D architecture the horizontal computation and the vertical computation are carried out in an intertwined manner. For example, in the visual cortex feed forward LaC computation from primary visual cortex to secondary and tertiary visual areas starts in layer 3 and terminates mainly in layer 4. Feed back LaC computation on the other hand typically originates in cells in layers 5 and 6 and terminate in layers 1, 2 and 6 [12]. Similarly LoCs also span across layers and not just remain confined within a single layer as initially appeared in Figure 1.

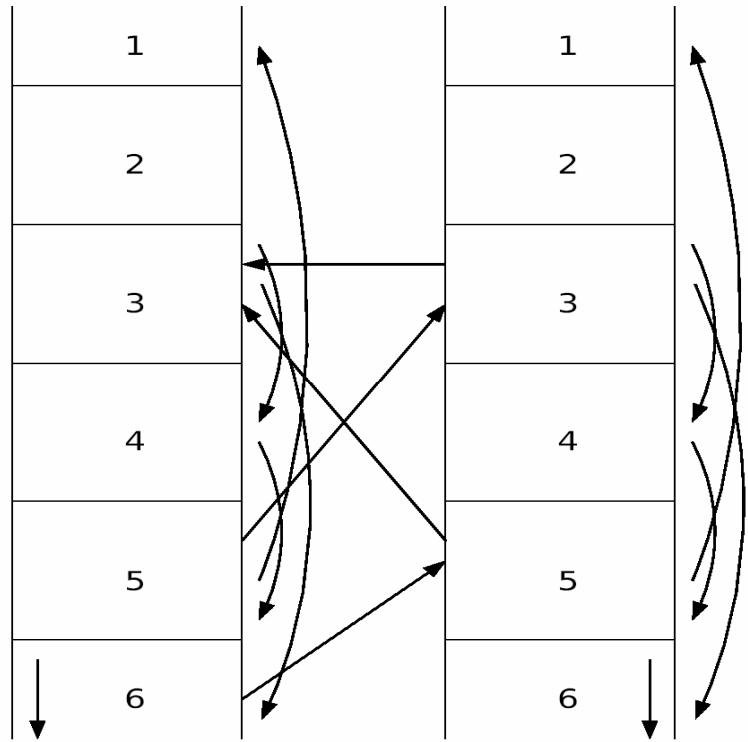

Figure 2. Vertical computation (VeC) in cortical columns, which signifies information processing across all the six layers of the cortex (drawn after Figure 17-9 of [12]). It should be viewed in conjunction with Figure 1 because this figure signifies the passage of horizontal and vertical lines of Figure 1 trough the layers of 1 through 6 in which neurons of each Brodmann's area are organized.

In order to appreciate the 3D architecture of cortical computation Figure 1 and Figure 2 are to be viewed together. Layers 2 though 6 of Figure 2 (when the columns are assembled together to form the whole cortex) are of the form of Figure 1. Layer 1 does not contain any neuron, only dendrites and therefore takes part in communication, but not in direct computation (in this paper it has been assumed that cortical computation is performed by firing neurons only). A neuron always remains fixed in a particular layer of Figure 2, but dendrites and axons as horizontal and vertical lines of Figure 1 traverse freely from layer to layer of Figure 2. We are now in a position to state the following

**Theorem 3.1:** At most 27.82% of the cortical lines would loop around for feed back information processing in the cortex. If output of the thalamus is taken as the input to the cortex and cortical feed back has to regulate the information flow through the thalamus then the same is true for the thalamocortical circuit as well.

**Proof:** The digraph of the cortex or the cortical circuit consisting of neurons and synapses start evolving in the womb and continues throughout the life through cortical rewiring. Due to the enormous packing density of neurons in the mammalian cortex (in the human cortex it is of the order of $10^4 \, neurons/mm^3$, because the total surface area of human cortex is $2500 \, cm^2$ [21], maximum thickness is 3.2 mm [22] and the number of neurons in it is of the order of $10^{10}$ [17]) the development of cortical lines through growth of new synapses may be modeled as a symmetric random walk in $R \times R \times Z_6$. Cortex has six distinct layers. If all the synapses are excitatory then the following argument assures that a cortical line will loop around with probability 0.2782.

In a symmetric random walk [8] let $u_n$ be the probability that the nth random step leads to the initial position. Let $f_n$ be the probability that the nth step takes the walk back to the initial position for the first time. Clearly, $f_0 = 0$ and $u_0 = 1$. Let $f = \sum_{n=0}^{\infty} f_n$. Since the walk can return to its starting point with a probability in $[0, 1], f \leq 1$. We want to prove that for a symmetric random walk in $R \times R \times Z_6$ $f = 0.2782$.

If the walk is to return to its starting point at the nth step, it can do so by reaching the starting point for the first time at the first step and then again revisiting it at the (n-1)th step after that. Or, it can do so by coming back to the starting point for the first time at the second step and then revisiting it after (n-2)th step. Or, in general coming back to the starting point at the rth step for the first time and then revisiting it in the (n-r)th step. So

$$u_n = \sum_{r=1}^{n} f_r u_{n-r}. \tag{3.1}$$

Let us define $U(s) = \sum_{i=0}^{\infty} u_i s^i$ and $F(s) = \sum_{i=1}^{\infty} f_i s^i$ as two generating functions. In the right hand side of (3.1) is the convolution $f_r * u_r$ with the generating function $F(s)U(s)$. However on the left of (3.1) $u_0$ is missing. So from (3.1) we get

$$U(s) = \frac{1}{1 - F(s)}. \tag{3.2}$$

$$\sum_{n=0}^{\infty} u_n = \frac{1}{1 - f}. \tag{3.3}$$

So in order to show that a symmetric random walk in $R \times R \times Z_6$ sooner or later forms a loop with probability $f$ we need to find $\sum_{n=0}^{\infty} u_n$.

If $R \times R \times Z_6$ is denoted by X, Y and Z coordinates then the return to the initial position will be possible if and only if number of steps in positive and negative X direction are equal and also the number of steps in positive and negative Y and Z directions are equal respectively. This implies $u_n = 0$ if $n$ is odd. Since the walk is symmetric the probability of moving to positive X direction = probability of moving to negative X direction = probability of moving to positive Y direction = probability of moving to negative Y direction = probability of moving to positive Z direction = probability of moving to negative Z direction = $\frac{1}{6}$. So

$$u_{2n} = \frac{1}{6^{2n}} \sum_{\substack{i+j+k=n \\ k \in \{0,\ldots,5\}}} \frac{(2n)!}{(i!)^2 (j!)^2 (k!)^2}.  \tag{3.4}$$

Applying Stirling's formula and the fact that $\sum_{k=0}^{n} \binom{n}{k}^2 = \binom{2n}{n}$, we get

$$\sum_{n=0}^{\infty} u_{2n} \approx u_0 + \frac{1}{\pi} \sum_{n=1}^{\infty} \left(\frac{2}{3}\right)^{2n} \frac{1}{n} + \frac{1}{4\pi} \sum_{n=1}^{\infty} \left(\frac{2}{3}\right)^{2n} n + \frac{1}{4 * 16\pi} \sum_{n=1}^{\infty} \left(\frac{2}{3}\right)^{2n} n(n-1)^2 +$$

$$\frac{1}{36 * 64\pi} \sum_{n=1}^{\infty} \left(\frac{2}{3}\right)^{2n} n(n-1)^2(n-2)^2 + \frac{1}{576 * 256\pi} \sum_{n=1}^{\infty} \left(\frac{2}{3}\right)^{2n} n(n-1)^2(n-2)^2(n-3)^2 +$$

$$\frac{1}{14400 * 1024\pi} \sum_{n=1}^{\infty} \left(\frac{2}{3}\right)^{2n} n(n-1)^2(n-2)^2(n-3)^2(n-4)^2. \tag{3.5}$$

(3.3) and (3.5) with the facts that $u_0 = 1$, and $u_n = 0$ whenever $n$ is odd, give $f = 0.2782$. It is enough to calculate for up to $n = 100$ for all the practical purposes, because after that all the terms in (3.5) become too insignificant. □

Since many of the cortical synapses are inhibitory the actual amount of cortical lines looping around would be less than 28%. This actually leaves a larger scope of choice for the loop formation ($^{100}C_K > {}^{100}C_{28}$, where $K < 28$) among the cortical paths and therefore formation of loops under the influence of genes and environment becomes an overwhelmingly important prerequisite for cortical development. Relatively low occurrence of loops in the cortex has also been supported anatomically (Fig. 4 in [25], where it has been observed – most connections between areas in the cortex appear to be reciprocal but within an area the distributions of the forward and backward components do not precisely coincide).

Apart from the random walk argument, which gives a way to mathematically prove the existence of cortical loops, arguments based on the notion of computational complexity also leads to existence of cortical loops for the 'efficient' processing of visual information [23], [24]. Visual search is a basic operation for any visual information processing. It has been shown in [23] that the stimulus driven search alone is an NP-complete problem and it has been argued in [24] that to make the visual search task tractable it must also have to be goal driven. This implies the existence of loops consisting of feed forward (for the stimulus driven part) and feed back (for the goal driven part) lines. It has been inferred in [23] and [24] without proof that the same holds true for other sensory information processing also.

One remarkable aspect of the contrast and amplify principle is clearly differentiating biological neural networks (BNN) from the artificial neural networks (ANN). In an ANN the boundary of a pattern is traced with the help of a finite number of points (determined by the number of input nodes) and an interpolation function (determined by the synaptic weights). On the other hand BNN traces a pattern from the interior with the help of contrast principle and amplify that for a definite recognition. If for example, a visual pattern is decomposed into topologically connected components, closely spaced retinal ganglion cells would receive light rays from a given component which will be carried through closely spaced LaC lines to the higher processing cortical areas. More generally the following

**Theorem 3.2:** The LaC, LoC and VeC processing of a visual pattern preserves the topological structure.

**Proof:** Photon is reaching on every cell in retina from a wide area of the visual scene (assuming it as a two dimensional frame, the argument in this proof is valid for single retinal processing only, it is not valid for simultaneous processing by two retinas). If $R_{i,j}$ is a retinal cell located on a concave spherical surface receiving photons from area $S_{i,j}$ on the Euclidean plane of the visual surface, we can call $S_{i,j}$ the receptive field of $R_{i,j}$. $i, j$ are integers. Clearly $S_{i+m,j+n} \cap S_{i,j} \neq \phi$ (the null set) when $m, n$ are small integers and $\Im = \bigcup_{i,j} S_{i,j}$, where $\Im$ signifies the whole visual scene. In fact each $S_{i,j}$ has a very vaguely defined boundary and therefore we can assume it to be an open set. Then the collection of $S_{i,j}$'s forms an open basis for the relative topology on $\Im$ as a subspace of the Euclidean plane. As soon as a class of $R_{i,j}$ is activated by the photons coming from $\Im$ a collection of LaC paths and along with them associated LoC and VeC paths are also activated.

Now consider the processing in a single LaC line. Inside the line the signal received from $S_{i,j}$ is being transmitted from one membrane patch to the next through neurotransmitter filled vesicles (from a neuronal membrane to a presynaptic membrane and then from the postsynaptic membrane to the next neuronal membrane and so on). Each of these membrane patches has again a very vaguely defined boundary. In fact if a neuron or a synapse is taken as enclosed by a smooth closed surface then they have no boundary at all! So each membrane patch can be taken as an open set. Then the LaC processing always maps an open set onto another open set. LoC and VeC processing are to support LaC processing only and they too process information in the same membrane to membrane manner. Therefore the whole information processing from each $S_{i,j}$ under the LOVE architecture is an open mapping. Note that each neuron is disjoint from any other neuron so is each synapse. But when two synapses $s_p$ and $s_q$ belonging to the same functional area in the cortex are processing information from the same $S_{i,j}$ the output of $s_p$ and $s_q$ also largely overlap. The same is the case for two neurons in the same functional area of the cortex. This means that although neurons and synapses are discrete entities outputs of their information processing as membrane devices together form a continuous sheet like structure just like the (two dimensional) visual scene itself.

Now different cortical regions process different aspects like motion, form, color, etc of $S_{i,j}$. So multiple sheet like copies of $S_{i,j}$ are made and they are processed in a distributed manner in different regions of the cortex for various aspects. Apparently it looks like destroying the one to one correspondence with $S_{i,j}$, but it does not. Because $S_{i,j}$ itself is a superposition of multiple copies, some are for form, some other are for color and if eye balls are to be moved to keep track of it then its various copies at different positions of the eye balls are to be processed for motion, etc. If this is accepted then there is of course a one to one correspondence between $S_{i,j}$ and the collection of its copies in the cortex.

At the time of bottom up processing very little information from each $S_{i,j}$, if at all, reaches up to the highest processing areas of the cortex. During the top down processing an even smaller subset is chosen for the final processing. Which information would be accepted for processing during the bottom up and top down pass is determined by genetic make up and past experience. But once the choice is made the processing is performed in loop (within the LOVE architecture) and this means if the information flow during the bottom up processing is denoted by the function $f$ then the flow during top down processing can be denoted by another function $g^{-1}$. We have seen that both of them are one to one correspondence and both are open mapping. So also is $g^{-1}f$.

To complete the proof we only need to show $g^{-1}f$ is continuous. Let $x, y \in \Im$ and $|x - y| < \varepsilon$, where $|.|$ denotes Euclidean norm and $\varepsilon > 0$. Either both of $x, y$ are in same $S_{i,j}$ for some $i, j$ or they belong to different $S_{i,j}$'s. In any case when $\varepsilon$ is small the signal from $x$ and $y$ are going to be processed either in a single LaC line or closely spaced cortical lines (both LaC, LoC and VeC) for each of the features (color, motion, form etc) processing. So there is a $\delta > 0$ such that $|g^{-1}f(x) - g^{-1}f(y)| < \delta$. This completes the proof. □

Unfortunately the above proof is not rigorous enough. But we can hope that with more precise knowledge about visual signal processing the proof can be made more rigorous, probably along the above line of arguments. The result is however unlikely to be true for the auditory signals i.e., signals from spatially disconnected sources may be mapped to a connected region of the cortex. Anyway Theorem 3.2 in conjunction with Theorem 2.1 asserts that in case of a visual pattern a topologically connected region within the pattern in the environment would be processed as a connected region in the brain by a collection of closely contiguous LaC lines i.e., a single LaC path.

So the pattern recognition by BNN is not by approximating the boundary contour (hyper-surface) with piecewise line (or hyper-plane) segments like in ANN, but simply by selecting the appropriate collections of LaC lines i.e., by choosing the right LaC paths. Although there are considerable theoretical ([23], [24]) evidence for the validity of the following Theorem 3.3, under the light of the above arguments it needs a rigorous proof which will be omitted in this paper.

**Theorem 3.3:** Visual pattern recognition in the BNNs is computationally more efficient (performed in linear time) than the same by the ANNs.

Cortical computation as a combination of **L**aC, **Lo**C and **Ve**C can be termed as LOVE and let us call the architecture shown in Figures 1 and 2 as the LOVE architecture.

**4. Memory and Learning**

Both long term and short term memory reside within the cortex as a 3D mesh of dendrites, neurons and axons i.e., in the LaC, LoC and VeC paths. A particular memory in a network depends on
      (1) the exact geometrical (or architectural) structure of the network, and
      (2) the exact level of neurotransmitters/neuromodulators in that network.
The above two conditions along with a third one,
      (3) the types of nerve cells involved,
are also the precise conditions for computation by the network. In the LOVE architecture the third condition has been taken care of by the functional areas of the cortex (say according to the classification of Brodmann), the second one is a functional aspect and falls outside the purview of the network architecture (functional stability has been dealt in [10], [13]). For the first condition it has been shown in the previous two sections how the cortex can be described in terms of the LOVE architecture.

(1) ensures the geometry of dendritic arbor and (2) ensures given an input how 'smoothly' it would be carried to the neuron through the arbor. Therefore (1) and (2) together ensure a very specific activity level for the neurons present in the network with respect to a given input. This will lead to recreation of a cognitive experience (created earlier by that particular input) of the network (memory recall). When the stimuli from the environment reach the thalamus they activate all the LaC paths they encounter. In fact if the inputs to the LaC paths (meant to process a particular type of sensory stimulus) at a particular instance are assumed to be organized on a two dimensional Euclidean plane the stimuli will draw a photographic image on that plane (the so called self organizing map [14]). In a geometric sense this is true not only for visual stimulus but also for all other forms of sensory stimuli. In this image each (noticed) feature of the external environment would have a unique representation in the brain. Organization of some of the features (if not all) will activate a set of LaC paths, which in turn will activate specific nerve cells through which a whole lot of other LaC paths will be activated as described in the previous sections (elaborated in Figure 3 and further described in the current section). Along with the LaC paths appropriate LoC and VeC paths will also be activated as described before. Very quick sequential activations of the neural networks in the brain like this may lead to recall of a whole experience from a fraction of the stimuli which took part in creating the experience initially.

**Definition 4.1:** Let $D$ be a data set. *Critical set learning* of $D$ is learning the whole of $D$ by a smallest subset $S$ of $D$. $S$ is called the *critical set*.

This means if $D$ can be learned by a chain of proper subsets $S_1 \subset S_2 \subset .... \subset D$ then $D$ can also be learned by $S = \bigcap_i S_i$, where $S$ is the critical set.

**Conjecture 4.1:** Given a LaC path and a learning by that path, almost always there is a minimum collection of LaC lines within that path, activating which after the learning will activate the whole LaC path involved in the initial learning. In other words cortical learning is almost always a critical set learning.

Learning by a LaC path signifies the neurons within that path will produce exactly the same spike trains in response to the same stimulus (this is a simplified assumption however, for even when a neuron is oscillating below threshold for spike initiation, it can still release neurotransmitter and shape the final circuit output [15]). 'Almost always' means here the collection of events for which the assertion does not hold has measure zero. Event here means a point in the pseudometric space $\left(R^{2r+1}, d\right)$ defined in [10], where $R^{2r+1}$ is the $2r+1$ dimensional real linear space and $d$ is the pseudometric defined on it by (3.5) in [10]. In everyday language this means in the vast majority of cases a previously known object is recognized by our brain even when the object is exposed partially. This is true as long as the signal reaching the sensory neurons are enough to activate the critical sets in the LaC paths which were used to learn the object initially. This issue was addressed by Hopfield from a different but equivalent point of view [9]. I shall return to a specific example of crtical set learning under the LOVE architecture later in this section.

But how that experience of learning got to be stored in the network in the first place? A succinct answer to this question can be given by quoting the *synaptic plasticity and memory hypothesis* as enunciated in [16], "Activity-dependent synaptic plasticity is induced at appropriate synapses during memory formation and is both necessary and sufficient for the information storage underlying the type of memory mediated by the brain area in which that plasticity is observed." To elaborate it under the LOVE architecture let us note that once a stimulus is presented before a sensory organ and if its signal is carried through a LaC path in the cortex it will propagate by the paired pulse facilitation (PPF) (a good description of PPF can be found in Chapter 13 of [17]). The duration of PPF is of the order of 100 ms. While some synaptic facilitation is induced in the network, consisting of LaC, LoC and VeC paths, after a single stimulus, the degree of facilitation increases with the number of stimuli. As the number and frequency of stimuli increase, another form of potentiation, *augmentation* is induced. Further stimulation brings into play a third form, termed *posttetanic potentiation* (PTP) [17]. The duration of augmentation is of the order of 10 s, while the duration of PTP is of the order of minutes. Since facilitation increases the probability of release of neurotransmitter the more number of LaC lines are activated the greater is the chance of reaching the signal up to the higher processing areas of the cortex such as the prefrontal cortex, where decisions are made or the hippocampus which mediates the storage of long term declarative memory or the

amygdala which mediates emotion. PPF, augmentation and PTP can be described by the following equation [17]

$$p(t) = p_0 + (p_f - p_0)\exp\left(\frac{-t}{\tau_f}\right), \tag{4.1}$$

where $p(t)$ is the probability of (neurotransmitter) release at time $t$ ($0 \leq t \leq \tau_f$), $p_0$ and $p_f$ are respectively probabilities of release before and after facilitation, $\tau_f$ is the characteristic decay time. (4.1) represents PPF when $\tau_f$ is of the order of 100 ms, augmentation when $\tau_f$ is of the order of 10 s, and PTP when $\tau_f$ is of the order of minutes. Note that (4.1) gives an experimental opportunity to compute $p_0$ and $p_f$ from $p(t)$ at different $t$. Putting the value of $p_0$ in

$$p_{A,B} = \sum_{j=k}^{r} \frac{r!}{j!(r-j)!} p_0^j (1-p_0)^{r-j}, \tag{4.2}$$

we get $p_{A,B}$ which is the association probability between $A$ and $B$ as described in the answer to question 2.1.

A LaC path $A$ activated by signals from environmental stimulus activates another LaC path $B$ in the cortex with probability $p_{A,B}$. If such activation happens in a LoC cycle (Figure 3) in higher cortical processing regions, consisting of excitatory feed forward synapses then several LaC paths may become active just by the PPF and/or augmentation and/or PTP generated in a single LaC path (say $A$) directly by the environmental stimulus (Note that in Figure 3 all this have been shown by LaC lines only, which needs to be generalized for LaC paths). In other words the cognitive computation in $A$ may invoke a whole bunch of other paths distributed over a large network in the cortex. Since a loop can self-sustain by propagation of feed forward and feed back signals even a transient stimulus to $A$, if strong enough, will be able to keep circulating the flow of PTP through the LaC, LoC and VeC paths (since VeC goes on in conjunction with LaC and LoC it will be kept tacit in most of the discussions) of a large network to keep it active even long after the cessation of the environmental stimulus. This whole activity manifests as *short term* or *working* memory. To this if activities of motor neurons are also added we get *cognition*.

For preserving long term memory anatomical change in the brain is necessary through formation of new synapses. Formation of new synapses needs synthesis of new proteins, which in turn needs activation of specific genes in the chromosome of the nerve cells. From a LOVE architectural point of view the significance of formation of new synapses to preserve long term memory can be summarized in Figure 3. In this figure the LaC lines (and hence also the LaC paths) have been classified into two distinct classes – one which takes input directly from the environment will be called *Type I LaC* and the other which does not take input directly from the environment will be called *Type II LaC*. The input taken through Type I LaCs are transmitted from lower processing areas of the brain (say thalamus) to the higher processing areas such as the hippocampal formation or the prefrontal cortex. Type II LaCs are connected to Type I LaCs by ordinary synaptic connections as well as through the LoC loops spanning across the higher processing regions as shown in Figure 3. Type I LaCs activate the appropriate LoC loops, which in turn activate the Type II LaCs. Activated Type II LaCs by self-sustained loop systems keep on processing the environmental stimulus even after the cessation of the stimulus itself in the environment. This in turn necessitates formation of new synapses between pairs of neurons from Type I LaCs and Type II LaCs (shown in light lines in Figure 3). Both the neurons are excitatory and synaptic joints are shown in light lines to indicate they are either in the formation stage or about to be strengthened by Hebbian rule. Formation of new synapses in the nervous system of *Aplysia* as a consequence of new learning has been stated in [4]. Formation of new synapses in response to bursts of synaptic stimulation in the mammalian cortex has been described in [18].

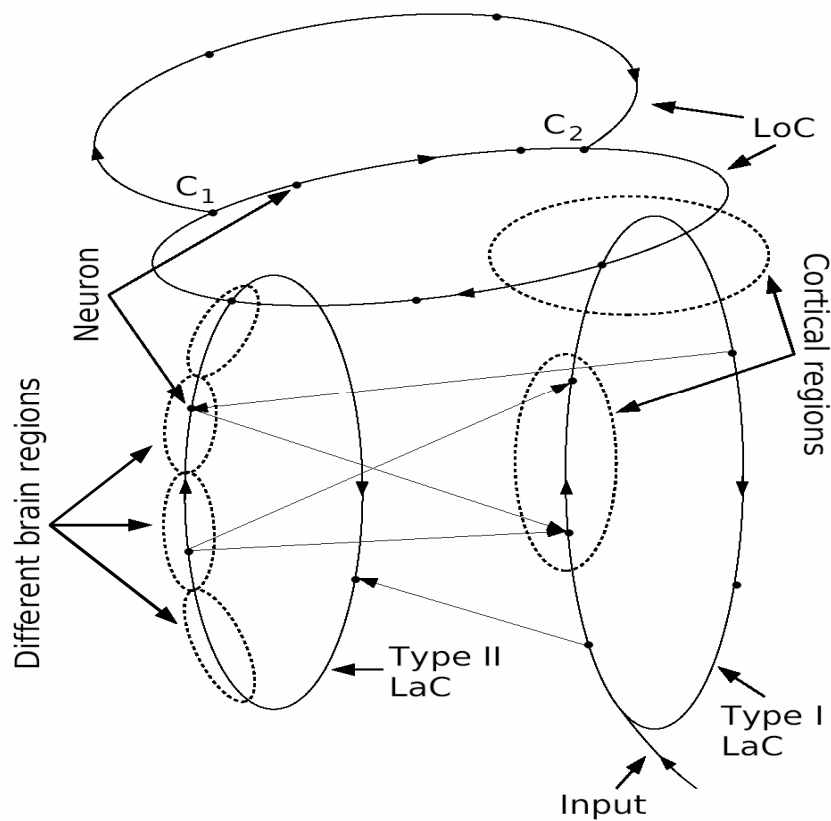

Figure 3. A typical interaction of a collection of looping LaC lines with a looping LoC line (another LoC line is not looping and therefore is not a cycle) spanning across some of the higher processing areas of the cortex, such as hippocampal formation and prefrontal cortex. The lighter lines indicate synaptic connections to be strengthened according to the Hebbian rule.

Type I LaCs are always dedicated to receive inputs directly from the environment and therefore must be set free for that task leaving the back ground processing of the already acquired information to the Type II LaCs. A Type I and a Type II LaC are connected through ordinary synapses as well as through the LoCs spanning through the higher processing regions of the cortex (such as hippocampus) as shown in Figure 3. When Type I LaCs activate Type II LaCs through the higher processing region LoCs the other synapses connecting the neurons of Type I LaCs with those of the Type II LaCs get strengthened by Hebbian rule, because the neurons on either end of each of them have been activated along with the respective LaC lines they are situated in. This needs synthesis of new proteins (for detail see ref. [4]) and may take hours to days to complete the formation of new synapses and augmenting their strength. Spontaneous firing of hippocampal pyramidal neurons may help to keep the LoCs active, which in turn keep the LaCs active. During this process back ground processing of the environmental stimuli keep continuing in the Type II LaCs leading to activation of specific genes, synthesis of new proteins and formation of new synapses. This is the key to converting the short term memory into the long term memory. Once this is done stimuli through Type I LaCs can activate the Type II LaCs without the involvement of higher processing region LoCs. Since hippocampus may play an important role to keep higher processing region LoCs active, this may be the reason why hippocampus is not so much needed to recall the already formed declarative memory as it is needed to form them initially.

Notice that Type I LaCs are to be made free for accepting subsequent inputs. When the new set of inputs arrive they too are likely to reach the Type II LaCs who are busy in processing the previous inputs. This means the Type II LaCs will have to be perturbed before the short term memory in them can be converted into long term memory. If this is true then formation of long term memory would be very difficult and for most of the stimuli no long term memory would have been preserved. This apparent contradiction is resolved through combinatorial assembly of LaC lines in a LaC path and the same for LoC lines in a LoC path. At a given time signal will be processed in a given LaC line in a particular manner. It would have a particular interpretation depending on which particular assembly of LaC and LoC lines it is being considered a part of. *This is the essence of the dynamic core hypothesis*. Some of these combinatorial assemblies of the LaC and the LoC lines within the LaC and the LoC paths respectively are genetically rewired, others are rewired through interactions with the environment after birth. Defective genetic rewiring may be a cause of mental retardation. The combinatorial assemblies of the LaC and the LoC lines within the LaC and the LoC paths respectively warranty a rather elaborate mathematical study, which would be taken up in future papers.

In Figure 4 an example of critical set learning within the LOVE architecture has been elaborated. Signal from a particular stimulus, say $S$, has been received and sent to cortex via the thalamus through the Type I LaC path shown in the Figure 4. Some features of $S$ may have appealed more to the individual cortex depending on the memory of past experiences and emotion (such as lips and eyes in a whole face) and accordingly those features have sent a stronger signal than the others and they use a subset of the LaC looping lines within the LaC path as shown in Figure 4. This subset has been called the critical set in this paper. When the signal reaches the hoppocampal LoCs through the critical aset of LaC looping lines, the signal is able to activate a set of pyramidal neurons there which are connected in a loop often with the ability that whenever only one presynaptic neuron fires, the postsynaptic one also fires. This in turn activates the hippocampal LoC looping path leading to activation of the Type II LaC path used to preserve the memory of $S$ (this memory may typically include some other features of $S$ along with the context and environment in which it was presented, albeit relatively faintly). The combination of connections among the presynaptic and postsynaptic neurons across the LaC looping lines of the LaC paths has not been shown in Figure 4. It is this connections which are crucial for activating the whole LaC path by making only a few neurons fire. The minimal input which can activate the appropriate LaC and LoC paths to recall the memory of $S$ can be termed as *critical input*.

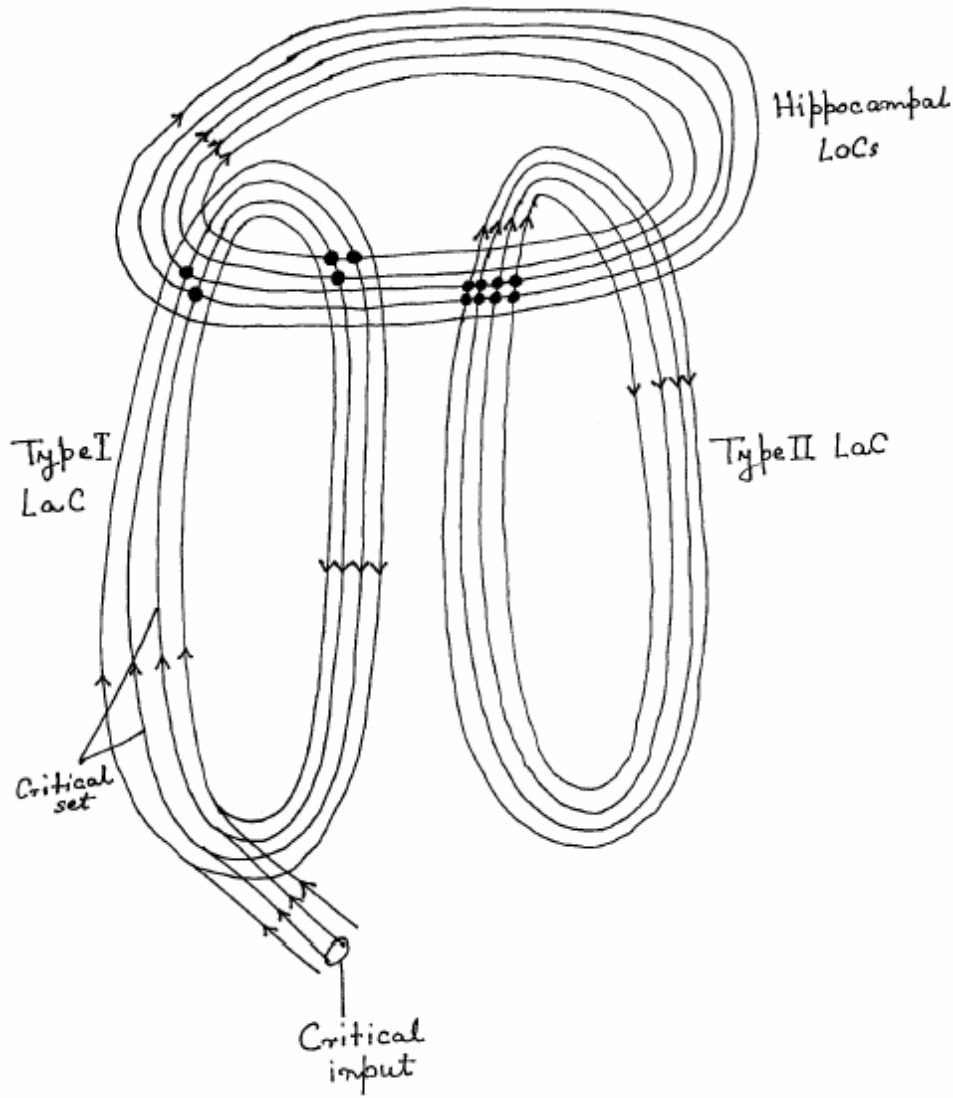

Figure 4. A critical set learning under the LOVE architecture. The whole Type I LaC path was activated by the stimulus when it was presented for the first time, but by the trick of cortical connections through new synapses (shown in thick black dots) only a subset of it is sufficient to recall the memory of the stimulus preserved in the shown Type II LaC.

It is clear that the critical set with respect to a stimulus will be very individual specific. It will also depend on the stage of development in life and therefore on time. The same critical set may not remain the critical set across the whole life of an individual. Since we are far from knowing the precise structure and function of cortical circuits we are not in a position to design an algorithm for determining a critical set in case of learning a stimulus by an individual at certain point of time in life. However in a graph theoretic model at much more elementary level it is possible to develop such algorithms. These algorithms may not be useful in neuroscience, but may find use in neural networks and artificial intelligence.

**Discussion and conclusion**

Although the brain computes as do the electronic computers their architectures are different. Unlike electronic computers a brain does not compute with a von Neumann type architecture. In this paper an architecture, christened as LOVE, for information processing by the cortical circuits has been proposed.

It is already well recognized that the brain functions are largely controlled by the architecture of the cortex. It is therefore imperative to understand the architecture of the cortex in order to understand the functions of the brain. Cortex on the other hand is a most complex structure in the universe, where neurons and synapses remain connected in a well structured network with enormous packing density. Also this structure is highly dynamic – anatomical changes may take place within hours. It has been experimentally and theoretically well established that the nervous systems process significant amount of information through feed back loops. In this paper cortex has been viewed as an assembly of mutually overlapping loops. In fact the loops are so overlapping that it is virtually impossible to have a unique decomposition of the cortex into loops.

In this scenario some scheme had to be adopted to decompose the cortex into loops, such that the logical operations remain intact yet the decomposition significantly simplifies the cortical architecture. Cortex is a three dimensional structure, which can be thought of as a sheet of paper with six layers. Then what can be more simple than a classical orthogonal type decomposition? The cortical loops should be decomposed along X, Y and Z axes, where the Z-axis can admit only six values – one for each cortical layer. The rule followed in this paper are (1) the loops which carry signals from thalamus or sensory neurons to the highest processing regions of the cortex and bring the output back are arranged along the X-axis and named lateral computing or LaC loops; (2) the loops which coordinate among the LaC loops across and within various cortical areas are to be arranged along the Y-axis and are called longitudinal computing or LoC loops; (3) the loops which coordinate among various layers of the cortex are arranged along the Z-axis and are called vertical computing or VeC loops. Note that this is only a convenient decomposition of big cortical loops, which have sub-loops of various sizes spanning within and across regions. LaC and LoC loops traverse through layers of the cortex and therefore VeC remains part of each and every LaC or LoC. Usually it is possible to decompose a given sufficiently large cortical loop into LaC, LoC and VeC parts in more than one different ways. If along with that loop an arbitrary collection of a sufficiently large number of other cortical loops overlapping with that one is also taken then it would be possible to decompose the collection into mutually overlapping LaC, LoC and VeC loops. The main hypothesis of this paper is – a typical task in the cortex is processed through a collection of mutually connected LaC, LoC and VeC loops. A probabilistic argument on line of symmetric random walks in three dimensions has been given to show that every cortical line has a certain probability to loop around.

The first major result of this paper is the contrast and amplify principle, where under the LOVE architecture it has been shown selection of specific LaC paths process specific features of the stimulus and during the process the signals from the stimulus get amplified by the LoC loops consisting of excitatory neurons from LaC lines within a LaC path and the feed forward synapses joining the neurons with one after another in the LoC loops. A LaC line may belong to more than one LaC paths. Signal processing in that line would have a particular meaning when it would be viewed as part of a particular LaC path. The same is true for LoC lines also. When this is true for a collection of closely spaced LaC or LoC lines it is true for a LaC or LoC path also. This dynamic changing of the utility of a line or a path from task to task is the architectural embodiment of the dynamic core hypothesis. This is true within a single task also if it can be decomposed into several sub-tasks and each of them is treated as a different task, which dynamically share a cortical 'core' consisting of LaC, LoC and VeC loops.

The second most significant result is the proof of existence of cortical loops with a probability of 0.2782 or less. This seems to be going against the importance of a loop based architecture, but it is not. Since only a small percentage of the cortical lines (and therefore also cortical paths) will form loop the choice of selection of those lines becomes much greater, for $^{100}C_K > {}^{100}C_{28}$, where $K < 28$. This will have profound implication on the development of our brain under the influence of learning. Let me hypothesize here that a baby is born with certain inbuilt information processing abilities in the brain and the cortical loops already formed in the womb take important role in those processing. Right from the birth the baby starts learning from the environment and in the process new synapses grow to make more and more new loops in the cortex. One implication of this may be we do not remember much of our life before five years of age. This happens because not enough cortical loops have developed to put a new percept associated with a large number of contexts (each feature of a context would need a separate cortical looping path to perceive according to the LOVE architecture). Lac of enough loops keeps the ability to perceive the number of contexts limited. Since a percept is not embedded in the memory of enough number of contexts it gets lost easily and quickly. From the arguments of section 4 it is clear that if a particular percept is embedded in a large number of contexts (each needing cortical looping paths to be stored) then it gets the chance to be refreshed from time to time through activating one or the other looping paths associated with one or the other context respectively. Otherwise synapses die more easily because of lack of activity and circuits get destroyed leading to loss of memory.

A couple of mathematical results have been stated in this paper without proof. This is to emphasize that the LOVE architecture described here, apart from being grounded on a strong neurophysiological foundation, is also conducive to rich theoretical study from mathematical and computational points of view. For example, Theorem 3.2 and Conjecture 4.1 are deep results and I guess the latter would take considerable effort to prove. However Theorem 3.3 may not be too difficult to establish from a computational complexity point of view.

In this paper a new paradigm of learning, called critical set learning, has been proposed, which seems to be the most natural in case of learning by a nervous system in general and the cortical learning in particular. The main idea is learning the whole only from a minimal set of partial information. For example, if we are told 'Einstein' most of us can readily recollect 'Albert Einstein'. It has been argued how LOVE architecture can implement such a scheme. The main result has been summarized in Conjecture 4.1, proving which will further establish the validity of the LOVE architecture.

The ideas presented in this paper has great potentiality for computer simulation. In future efforts will be made to develop an open source software (probably named 'cortex') which will simulate aspects of cortical computation within the setup of the LOVE architecture described in this paper. In this software a neuron will be represented as a spike train or more precisely as a vector consisting of the Fourier coefficients of a spike train over a period of time (as described in [10], [13]). Generally the spike train will keep changing from one period to the next, which will signify changing input to the neuron. Neurons will form loops and loops will form the cortical circuit according to the LOVE architecture (of course loops will have specific orientations, not every two neurons will belong to a loop). The algorithm controlling the functions of the circuits will be as described in [10] and [13]. The functional aspect described in [10] and [13] take into account both past experiences and the level of impulse (the simplest form of emotion). Impulse or the simplest form of emotion has been quantified in [10] and [13] with the help of a mathematical function. This brings the motivation under the purview of computation. I am not aware of any other attempt so far to quantify emotion.

## Acknowledgement

The author likes to acknowledge Chitta Ranjan Das for drawing the figures. Critical comments by two anonymous reviewers have greatly helped to improve the paper. A postdoctoral fellowship from the Institute of Mathematical Sciences (funded by the Department of Atomic Energy of the Government of India) under which the current research has been carried out is being acknowledged.

## References

1. M. Moscovitch, R. S. Rosenbaum, A. Gilboa, D. R. Addis, R. Westmacott, C. Grady, M. P. McAndrews, B. Lavine, S. Black, G. Winocur and L. Nadel, "Functional neuroanatomy of remote episodic, semantic and spatial memory: a unified account based on multiple trace theory," *J. Anat.*, vol. 207, pp. 35–66, 2005.
2. L. G. Valiant, "Cognitive computation," *Proc. 36th FOCS*, pp. 2–3, IEEE Comp. Soc., CA, 1995.
3. L. G. Valiant, "A neuroidal architecture for cognitive computation," *JACM*, vol. 47(5), pp. 854–882, 2000.
4. E. R. Kandel, "The molecular biology of memory storage: a dialogue between genes and synapses," *Science*, vol. 294, pp. 1030–1038, 2001.
5. G. Edelman and G. Tononi, *A Universe of Consciousness: How Matter Becomes Imagination*, Basic Books, 2001.
6. G. Edelman and G. Tononi, "Consciousness," *Science*, vol. 282, pp. 1846–1851, 1998.
7. A. K. Engel, P. Fries and W. Singer, "Dynamic predictions: oscillations and synchrony in top-down processing," *Nature Rev. Neurosci.*, vol. 2, pp. 704–716, 2001.

8. W. Feller, *An Introduction to Probability Theory and Its Applications*, vol. 1, 3rd ed., John Wiley & Sons, New York, 1968.
9. J. J. Hopfield, "Neural networks and physical systems with emergent collective computational abilities," *Proc. Natl. Acad. Sc., USA*, vol. 79, pp. 2254–2258, 1982.
10. K. Majumdar, "A structural and a functional aspect of stable information processing by the brain," to appear in *Cognitive Neurodynamics*, also available at http://arxiv.org/ftp/q-bio/papers/0701/0701034.pdf, 2007.
11. L. G. Valiant, "Memorization and association on a realistic neural model," *Neural Computation*, vol. 17(3), pp. 527–555, 2005.
12. E. R. Kandel, J. H. Schwartz and T. M. Jessell, *Principles of Neural Science*, 4th ed., McGraw Hill, 2000.
13. K. Majumdar, "Behavioral response to strong aversive stimuli: A neurodynamical model," http://arxiv.org/ftp/arxiv/papers/0704/0704.0648.pdf, (submitted to *Cognitive Neurodynamics*), 2007.
14. T. Kohonen, "The self organizing map," *Proc. IEEE*, vol. 78, pp. 1464–1480, 1990.
15. R. M. Harris-Warrick, and E. Marder, "Modulation of neural networks for behavior," *Annu. Rev. Neurosci.*, vol. 14, pp. 39–57, 1991.
16. S. J. Martin, P. D. Grimwood and R. G. M. Morris, "Synaptic plasticity and memory: an evaluation of hypothesis," *Annu. Rev. Neurosci.*, vol. 23, pp. 649–711, 2000.
17. C. Koch, *Biophysics of Computation: Information Processing in Single Neurons*, Oxford University Press, New York, 1999.
18. J. T. Trachtenberg, B. E. Chen, G. H. Knott, G. Feng, J. R. Sanes, E. Welker and K. Svoboda, "Long-term in vivo imaging of experience-dependent synaptic plasticity in adult cortex," *Nature*, vol. 420, pp. 788–794, 2002.
19. D. Mumford, "On the computational architecture of the neocortex: I. The role of the thalamo-cortical loop," *Biol. Cybern.*, vol. 65, pp. 135–145, 1991.
20. D. Mumford, "On the computational architecture of the neocortex: II. The role of cortico-cortical loops," *Biol. Cybern.*, vol. 66, pp. 241–251, 1992.
21. A. Peters and E. G. Jones, *The Cerebral Cortex*, vol. 1, Plenum Press, New York, 1984.
22. B. Fischl and A. M. Dale, "Measuring the thickness of the human cerebral cortex from magnetic resonance imaging," *PNAS, USA*, vol. 97(20), pp. 11050–5, 2000.
23. J. K. Tsotsos, "The complexity of perceptual search tasks," *Proc. Int. Jt. Conf. AI*, Detroit, USA, August 1989.
24. J. K. Tsotsos, "Analyzing vision at the complexity level," *Behavioral and Brain Sciences*, vol. 13, pp. 423–469, 1990.
25. S. Zeki and S. Shipp, "The functional logic of cortical connections," *Nature*, vol. 335, pp. 311–317, 1988.